\documentclass[prd,twocolumn]{revtex4}
\usepackage{bm}
\usepackage{epsfig}
\usepackage{amsmath}
\newcommand{\be}{\begin{equation}}
\newcommand{\ee}{\end{equation}}
\newcommand{\ba}{\begin{eqnarray}}
\newcommand{\ea}{\end{eqnarray}}
\newcommand{\bes}{\begin{subequations}}
\newcommand{\ees}{\end{subequations}}

\newcommand{\bi}{\begin{itemize}}
\newcommand{\ei}{\end{itemize}}

\begin{document}
\title{Shadow vacuum alignment and dark energy}
\author{P.Q. Hung}
\email[]{pqh@virginia.edu}
\affiliation{Dept. of Physics, University of Virginia, \\
382 McCormick Road, P. O. Box 400714, Charlottesville, Virginia 22904-4714, 
USA}
\date{\today}
\begin{abstract}
In a recent model of dark energy (with several
phenomenological consequences),
the universe is assumed to be trapped in a false vacuum with 
an energy density of
the order of $(10^{-3}\,eV)^4$, mimicking the
presently successful $\Lambda CDM$ scenario. This involves
a new gauge group $SU(2)_Z$, {\em the shadow sector}, which becomes strong 
at a scale $\Lambda_Z \sim 10^{-3}\,eV$. The model is described by
the $SU(2)_Z$ instanton-induced potential of an 
axion-like scalar field, $a_Z$, with two degenerate vacuua.
The false (metastable) vacuum appears as a result of an phenomenological
(ad-hoc) soft breaking term linear in $a_Z$ which explicitely breaks that degeneracy.
In this paper, we discuss
a possible dynamical origin for this soft breaking term as coming from the
alignment of the vacuum along
a direction in which the condensate of the shadow fermions,
$\langle \bar{\psi}^{(Z)}_{i}\,i\,\gamma_{5} \psi^{(Z)}_{i} \rangle$ 
which breaks spontaneously both $P$ and $CP$, is non-vanishing. The present universe
lives in a vacuum which violates both
$P$ and $CP$ in the shadow $SU(2)_Z$ sector!
\end{abstract}
\maketitle

The nature of the dark energy responsible for the present acceleration
of the universe is one of the most profound mysteries of modern
cosmology.The most recent data \cite{cosmodata} was found to be consistent
a picture of the universe- $\Lambda CDM$-which is dominated by a 
cosmological constant. The dark energy density is approximately
$(10^{-3}\,eV)^4$. Where does this number come from? Does an energy
$\sim 10^{-3}\,eV$ signal a {\em new} scale of physics? What would
be its origin? If it were a new scale, how does an energy density
$\sim (10^{-3}\,eV)^4$ arise? Can one derive it from a model if
there is one?

In \cite{hung1}, \cite{hung2}, a model was proposed 
in which the present universe is
trapped in a false vacuum with an energy density being approximately
$(10^{-3}\,eV)^4$. As explained in detail in \cite{hung2}, 
the acceleration is driven 
by an ``axion-like'' scalar field $a_Z$ whose potential is induced
by instanton effects of a new gauge group $SU(2)_Z$ (the shadow sector)
which grows strong
at $\Lambda_Z \sim 10^{-3}\,eV$. (As mentioned in \cite{hung2} and
discussed in detail in \cite{hung3}, 
this new gauge group $SU(2)_Z$ can be seen to come 
from the breaking $E_6$ into
$SU(2)_Z \otimes SU(6)$, where $SU(6)$ can, as one possible
scenario, first break down
to $SU(3)_c \otimes SU(3)_L \otimes U(1)$ and then to
$SU(3)_c \otimes SU(2)_L \otimes U(1)_Y$.) This scenario contains
a number of consequences \cite{hung2},
\cite{hung4}, beside providing a model for the dark energy: 
candidates for the cold dark matter in the form of fermions which
are triplets under $SU(2)_Z$ and singlets under the Standard Model (SM),
namely $\psi^{(Z)}_{(L,R),i}$ with $i=1,2$; a new mechanism for
leptogenesis in which it is the decay of a messenger scalar field
which carry quantum numbers of both $SU(2)_Z$ and the SM which creates
a net SM lepton number; and finally the possibility of detection
of the messenger field itself at future colliders such as the LHC,
as well as possible signals of the CDM candidates $\psi^{(Z)}_{(L,R),i}$
as missing energy in the decay of the messenger field.

In the model for dark energy expounded in \cite{hung2}, the potential
for $a_Z$ (which is the imaginary part of a complex scalar field) arises
due to $SU(2)_Z$ instanton effects as the $SU(2)_Z$ coupling grows large
at energies close to $\Lambda_Z \sim 10^{-3}\,eV$. As described
in \cite{hung2}, this potential has two degenerate vacuua because
of a remaining unbroken $Z(2)$ (for two flavors)
symmetry of the global $U(1)_{A}^{(Z)}$ symmetry which is present
in the model. This
degeneracy is lifted by having a phenomenological soft-breaking term linear in $a_Z$.
Because of this term, there is now a false vacuum at $a_Z \neq 0$,
with an energy density of the order of $(10^{-3}\,eV)^4$. It was proposed
that the present universe is trapped in this false vacuum 
with an equation of state  $w \approx -1$ and thus the scenario
basically mimics $\Lambda CDM$. The ages of various epochs as well as the
time it would take to exit from the false vacuum
to the true vacuum at $a_Z=0$ was estimated in
\cite{hung2}. 

While the form of the axion-like potential has been well studied in
the context of the Peccei-Quinn axion \cite{peccei}, the 
soft-breaking term which lifts
the vacuum degeneracy is slightly ad-hoc \cite{sikivie}. In Ref.
\cite{sikivie}, constraints were put on a parameter
which appears in that term in order to respect the bound on
the neutron electric dipole moment. 
Since a similar term was postulated in
\cite{hung1} and \cite{hung2}, one might wish to 
know more about its possible origin. Can one derive it
{\em dynamically} within the framework of the model
proposed in \cite{hung1} and \cite{hung2}?
The purpose of this note is to propose a mechanism to generate
this desired soft-breaking term in the context of the dark energy model
described in \cite{hung1} and \cite{hung2}. The upshot
of this mechanism is the interesting notion that the shadow
$SU(2)_Z$  false vacuum in which the universe is trapped is
$P$ and $CP$ odd! The vacuum energy density which characterizes the
dark energy, namely $\sim (10^{-3}\,eV)^4$, comes from a condensate
which breaks spontaneously both $P$ and $CP$ in the shadow sector.



In what follows we will only present the particle content given
in \cite{hung2} that is relevant to the present paper. 
In addition to the SM particles, our model
contains two fermions, $\psi^{(Z)}_{1,2}$, which are 
triplets under $SU(2)_Z$ and singlets
under the SM, one complex scalar field $\phi_Z$ plus two scalar
messenger scalar fields which are relevant to issues such as 
leptogenesis but are not needed here.
(The motivation for the aforementioned particle content of the model
is explained in detail in \cite{hung2}.)

The key ingredients to the dark energy model of \cite{hung2} are characterized
by the following global symmetry and interaction Lagrangians under
$G_{SM} \otimes SU(2)_Z$:
\ba
\label{lagrangian}
{\cal L}&=& {\cal L}_{SM} + {\cal L}^{Z}_{kin} + {\cal L}_{yuk}
+ {\cal L}_{CP} -V(|\tilde{\bm{\varphi}}^{(Z)}|^2
\; {\rm or}\; |\varphi^{(Z)}|^2)
\nonumber \\
& &-V(|\phi_{Z}|^2)\,,
\ea
\ba
\label{Zkin}
{\cal L}^{Z}_{kin}& =& -\frac{1}{4} {\bf G}_{\mu\nu}^{(Z)}.
{\bf G}^{(Z),\mu\nu} + (\sum_{i} \frac{1}{2} (D_{\mu}\,
\tilde{\bm{\varphi}}_{i}^{(Z)})^{\dag}.(D^{\mu}\,
\tilde{\bm{\varphi}}_{i}^{(Z)})\,   \nonumber \\
& & + \sum_{i} i \bar{\psi}^{(Z)}_{(L,R),i}\not\!D
\psi^{(Z)}_{(L,R),i}\,,
\ea
\ba
\label{yuk}
{\cal L}_{yuk}&= & \sum_{i}\,\sum_{m}( g_{\tilde{\varphi}_{1}\,m}^{i}\,
\bar{l}_{L}^{m}\,
\tilde{\bm{\varphi}}_{1}^{(Z)}\,\psi^{(Z)}_{i,R}+ 
g_{\tilde{\varphi}_{2}\,m}^{i}\,\bar{l}_{L}^{m}\,
\tilde{\bm{\varphi}}_{2}^{(Z)}\,\psi^{(Z)}_{i,R}) \nonumber \\
&& +\sum_{i} K_{i} \, \bar{\psi}^{(Z)}_{L,i}\,\psi^{(Z)}_{R,i}\,\phi_{Z}
+ h.c. \,,
\ea
\be
\label{cp}
{\cal L}_{CP} = \frac{\theta_Z}{32\,\pi^2} {\bf G}_{\mu\nu}^{(Z)}.
\tilde{{\bf G}}^{(Z),\mu\nu}\,,
\ee
where $G_{SM}$ is the SM gauge group, ${\cal L}_{SM}$ is the well-known SM Lagangian 
which do not need to be explicitely written here
and $\tilde{{\bf G}}^{(Z),\mu\nu}=
\frac{1}{2}\,\epsilon^{\mu \nu \lambda \rho}\,
{\bf G}_{\lambda \rho}^{(Z)}$. In \ref{yuk}, $i=1,2$ for
two $\psi^{(Z)}$ and $m=1,2,3$ for three families. 

As discussed in \cite{hung2}, our model contains a $U(1)_{A}^{(Z)}$ global symmetry.
Eqs.(\ref{yuk},\ref{cp}) are invariant
under the following $U(1)_{A}^{(Z)}$ phase transformation:
\bes
\be
\label{phase0}
\psi^{(Z)}_{i} \rightarrow e^{i\alpha \gamma_{5}}\,\psi^{(Z)}_{i}\,,
\ee
\be
\label{phase1}
\psi^{(Z)}_{L,i} \rightarrow e^{-i\alpha}\,\psi^{(Z)}_{L,i} \,,
\ee
\be
\label{phase2}
\psi^{(Z)}_{R,i} \rightarrow e^{i\alpha}\,\psi^{(Z)}_{R,i} \,,
\ee
\be
\label{phase3}
\phi_{Z} \rightarrow e^{-2i\alpha}\,\phi_{Z} \,,
\ee
\be
\label{phase4}
\theta_Z \rightarrow \theta_Z - 4\,\alpha \,,
\ee
\be
\label{phase5}
l_{L}^{m} \rightarrow e^{i\alpha}\,l_{L}^{m}\,,
\ee
\be
\label{phase6}
\tilde{\bm{\varphi}}_{i}^{(Z)} \rightarrow \tilde{\bm{\varphi}}_{i}^{(Z)}\,.
\ee
\ees
Since ${\cal L}_{SM}$ contains Yukawa couplings between the SM leptons
to the SM Higgs fiels $\phi_{SM}$ of the form $\bar{l}_{L}^{m}\,
\phi_{SM}\,l_{R}^{n}$ (and also $\bar{l}_{L}^{m}\,
\tilde{\phi}_{SM}\,\nu_{R}^{n}$ for the neutral leptons), 
where $l_{R}^{n}$ ($\nu_{R}^{n}$) denotes the charged (neutral)
right-handed leptons, it will be invariant under the above
$U(1)_{A}^{(Z)}$ global symmetry provided
\be
\label{phaser}
l_{R}^{m}\; (\nu_{R}^{m}) \rightarrow e^{i\alpha}\,l_{R}^{m}\;
(\nu_{R}^{m})\,,
\ee
when we use the transformation (\ref{phase5}). All other SM particles
are unchanged under $U(1)_{A}^{(Z)}$.

Similar to the Peccei-Quinn model \cite{peccei} in QCD, the above transformations
ensure that the Lagrangian is {\em explicitely} P and CP-conserving.
One can rotate $\theta_Z$ to zero and make the Yukawa couplings of
the shadow fermions real. However, unlike QCD, there is no reason
to suspect that there cannot be spontaneous P and CP violation in
the shadow sector. In fact, this interesting possibility will
be the focus of this paper.

There are several interesting features of the $U(1)_{A}^{(Z)}$ symmetry which
are summarized below. (Further details can be found in \cite{hung2}.)
\bi

\item The vacuum expectation value of $\phi_{Z}$ gives $\psi^{(Z)}_{i}$ their
masses: $|K_i| v_Z$, where $\phi_{Z} = v_{Z}\,\exp(ia_Z/v_{Z}) + \sigma_Z$
with $\langle \sigma_Z \rangle =0$ and $\langle a_Z \rangle =0$. It
was argued in \cite{lowinf} that a ``low-scale'' inflationary scenario
favours $v_Z \sim 10^{9}\,GeV$ and that $|K_i| \sim 10^{-7}$ in order
for $m_{\psi^{(Z)}_{i}} \sim O(100-200\,GeV)$.

\item The imaginary part of $\phi_{Z}$, namely $a_Z$, is the ``axion'' of
$SU(2)_Z$. Its $SU(2)_Z$ instanton-induced potential at zero
temperature is  given by
\be
\label{potential}
V(a_Z) = \Lambda_Z^4[1-\,\cos\frac{a_Z}{v_Z}] \,.
\ee
There is a remaining $Z(2)$ symmetry resulting in two degenerate vacuua. 
Such degeneracy is well-known
in the Peccei-Quinn axion potential as it has been noted by \cite{sikivie}.
Furthermore, \cite{sikivie} also pointed out that, because of the
$Z(N)$ degeneracy, the Peccei-Quinn axion
is incompatible with standard cosmology and proposed a soft-breaking term
linear in the axion field to lift this degeneracy. A similar proposal
was made for the $SU(2)_Z$ ``axion'' in \cite{hung2} where a term
of the form 
\be
\label{softbreak}
\Lambda^{4}(\frac{a_Z}{2\,\pi\,v_Z}) \,,
\ee
was added to (\ref{potential}) in order to lift the degeneracy of the two vacuua.

\item The scenario of the dark energy presented in \cite{hung2}
was based on Eqs. (\ref{potential},\ref{softbreak}) where
the present accelerating universe is assumed to be trapped in a false vacuum at
$a_Z = 2\,\pi\,v_Z$ with an energy density $\sim \Lambda^4$. Various
cosmological quantities were computed within the framework of the model,
including the (extremely large) exit time to the true vacuum at $a_Z=0$.

\item The pseudo Nambu-Goldstone (PNG) boson $a_Z$ acquires a mass
because $U(1)_{A}^{(Z)}$ is explicitely broken by $SU(2)_Z$ instantons, in
a very similar fashion to the PQ axion. This mass was computed in \cite{hung2}
to be $m_{a_Z}^2 = \frac{2\,\sum_{i} |K_{i}|\, \mu_{i}^3}{v_Z}$, where
$\mu_{i}^3$ is the shadow fermion chiral condensate value. (It is
estimated to be much less than $10^{-10}\,eV$.) What is interesting and most important
for this paper is the interaction of $a_Z$ with the $SU(2)_Z$ fermions $\psi^{(Z)}_{i}$,
$i=1,2$. From Eq. (\ref{yuk}), one obtains the following interaction
\be
\label{axionint}
{\cal L}_{a_Z} =  (\sum_{i} (\frac{m_{\psi^{(Z)}_{i}}}{v_Z})
\bar{\psi}^{(Z)}_{i}\,i \, \gamma_{5} \psi^{(Z)}_{i})\,a_Z\,.
\ee
Eq. (\ref{axionint}) can be interesting in studying the interaction of
$a_Z$ with the CDM candidates $\psi^{(Z)}_{i}$ of our model. However,
we will focus on this equation from another perspective, that of the
origin of the soft breaking term itself. The question we would
like to ask is the following: Could the soft breaking term (\ref{softbreak})
arises from a {\em non-vanishing} vacuum expectation value for
$\langle i\bar{\psi}^{(Z)}_{i}\,\gamma_{5} \psi^{(Z)}_{i} \rangle$ in
Eq. (\ref{axionint}) which breaks spontaneously P and CP?

\ei

In what follows, we will present a scenario in which the aforementioned
soft breaking term, Eq. (\ref{softbreak}), arises {\em dynamically}. 
It turns out that, in our model, the mass regime is of an extremely
convenient type for the study of condensates, namely
that of a heavy fermion, $m_{\psi^{(Z)}_{i}} \sim O(100-200\,GeV)
\gg \Lambda_{Z}$, similar to the treatment of heavy quark
condensates in QCD sum rules of Shifman, Vainshtein and Zakharov
\cite{svz}. In QCD where quarks belong to the fundametal representations, 
one can derive a relationship between the
heavy quark ($m_{Q} \gg \Lambda_{QCD}$) condensate
$\langle \bar{Q} Q \rangle$ and the gluon
condensate $\langle G_{\mu \nu}^{(QCD)}
G^{\mu \nu,(QCD)}\rangle$, namely $\langle \bar{Q} Q \rangle =
-\frac{2}{3\,m_{Q}}\frac{\alpha_S}{8\pi}\langle G_{\mu \nu}^{(QCD)}
G^{\mu \nu,(QCD)}\rangle$ where higher orders in $m_{Q}^{-1}$ are
neglected, especially when $m_{Q} \gg \Lambda_{QCD}$.

In our model of dark energy, $SU(2)_Z$ grows strong at
$\Lambda_Z \sim 10^{-3}\,eV$ and the $SU(2)_Z$ fermions
$\psi^{(Z)}_{i}$ have mass $m_{\psi^{(Z)}_{i}} \sim O(100-200\,GeV)
\gg \Lambda_{Z}$ in order to be CDM candidates as discussed in \cite{hung2}.
In consequence, the starting point of our discussion is
\be
\label{starting}
m_{\psi^{(Z)}_{i}} \sim O(100-200\,GeV) \gg \Lambda_{Z} \sim 
10^{-3}\,eV \,.
\ee
We are particularly interested in the pseudo-scalar condensates
of $\psi^{(Z)}_{i}$, namely 
$\langle i\bar{\psi}^{(Z)}_{i}\,\gamma_{5} \psi^{(Z)}_{i} \rangle$.
How is it related to $\Lambda_Z$?
In QCD with one extremely heavy quark, i.e. $m_{Q} \gg \Lambda_{QCD}$,
there is a relationship relating $\langle \bar{Q} \,i\,
\gamma_{5}\,Q\, \rangle$ to $\langle G_{\mu \nu}^{a}
\tilde{G}^{\mu \nu,a} \rangle$, where $a$ is an $SU(3)_c$ color index and
$\tilde{G}^{\mu \nu,a} = \frac{1}{2}\,\epsilon^{\mu \nu \lambda \rho}
G_{\lambda \rho}^{a}$. It is given by \cite{halperin}
\be
\label{ggtilde}
M_{Q}\, \langle \bar{Q} \,i\,\gamma_{5}\,Q\,  \rangle= 
-\langle \frac{\alpha_s}{4\,\pi}\,G_{\mu \nu}^{a}\,\tilde{G}^{\mu \nu,a} 
\rangle \,.
\ee 

An equivalent expression in our case will have to take into account the
normalization of the generators for $\psi^{(Z)}_{i}$ as well as its total 
number. One obtains
\begin{eqnarray}
\label{ggshadow}
\sum_{i}\,m_{\psi^{(Z)}_{i}}\,\langle i\bar{\psi}^{(Z)}_{i}\,\gamma_{5} 
\psi^{(Z)}_{i} \rangle &=& -n_{\psi^{(Z)}_{i}}\,(Tr\,T_{\psi^{(Z)}_{i}}^{2})
\nonumber \\
&& \times \langle\frac{\alpha_Z}{2\,\pi}\,{\bf G}_{\mu \nu}^{(Z)}\,
\tilde{{\bf G}}^{\mu \nu,(Z)}\,\rangle \,,  \nonumber \\
\end{eqnarray}
where $\alpha_Z = g_{Z}^2/4\,\pi$, $n_{\psi^{(Z)}_{i}}$ is the number of 
heavy fermions and
$Tr\,T_{\psi^{(Z)}_{i}}^{2}$ is the normalization of the generator
applying to the heavy fermion representation.
Notice that (\ref{ggshadow}) reduces to (\ref{ggtilde}) for
one heavy quark belonging to the fundamental representation.
In our case, with $\psi^{(Z)}_{i}$ belonging to the adjoint representation
of $SU(2)_Z$, $Tr\,T_{\psi^{(Z)}_{i}}^{2} = 2$ and $n_{\psi^{(Z)}_{i}} =2$,
we finally obtain
\be
\label{ggshadow2}
\sum_{i}\,m_{\psi^{(Z)}_{i}}\,\langle \bar{\psi}^{(Z)}_{i}\,i\,\gamma_{5} 
\psi^{(Z)}_{i} \rangle = -\langle\frac{2\,\alpha_Z}{\pi}\,{\bf G}_{\mu \nu}^{(Z)}\,
\tilde{{\bf G}}^{\mu \nu,(Z)}\,\rangle \,.
\ee

With (\ref{ggshadow2}), one can now obtain from Eq. (\ref{axionint}) the
following soft-breaking term
\be
\label{soft}
V_B = -\langle \Theta_Z |\frac{2\,\alpha_Z}{\pi}\,{\bf G}_{\mu \nu}^{(Z)}\,
\tilde{{\bf G}}^{\mu \nu,(Z)}\, | \Theta_Z \rangle \,(\frac{a_Z}{v_Z}) \,,
\ee
where we characterize the false vacuum by an angle $\Theta_Z$.
Two remarks are in order here. First, Eq. (\ref{soft}) represents the
term that lifts the degeneracy if  
$\langle \Theta_Z |\frac{2\,\alpha_Z}{\pi}\,{\bf G}_{\mu \nu}^{(Z)}\,
\tilde{{\bf G}}^{\mu \nu,(Z)}\,| \Theta_Z \rangle \neq 0$.
Second, we notice that 
$\langle \Theta_Z |\bar{\psi}^{(Z)}_{i}\,i\,\gamma_{5} \psi^{(Z)}_{i} 
| \Theta_Z \rangle$
and $\langle \Theta_Z | \frac{2\,\alpha_Z}{\pi}\,{\bf G}_{\mu \nu}^{(Z)}\,
\tilde{{\bf G}}^{\mu \nu,(Z)}\,| \Theta_Z \rangle$ breaks 
spontaneously both P and CP. (The Lagrangian is {\em explicitely}
P and CP-conserving as explained in the beginning.)
Therefore, the false vacuum of our scenario is one that violates P and CP in the
shadow sector. In order to proceed, one would like to know what
$\langle \Theta_Z | \frac{2\,\alpha_Z}{\pi}\,{\bf G}_{\mu \nu}^{(Z)}\,
\tilde{{\bf G}}^{\mu \nu,(Z)}\,| \Theta_Z \rangle$ might be.

In a similar fashion to \cite{huang}, let us
define the scalar and pseudoscalar condensates as follows
\bes
\label{thetaz}
\be
\sum_{i}\,m_{\psi^{(Z)}_{i}}\,\langle \Theta_Z | \bar{\psi}^{(Z)}_{i}\,
\psi^{(Z)}_{i}|\Theta_Z \rangle = -C\, \cos \Theta_Z \,,
\ee
\be
\sum_{i}\,m_{\psi^{(Z)}_{i}}\,\langle \Theta_Z |\bar{\psi}^{(Z)}_{i}\,i\,\gamma_{5} 
\psi^{(Z)}_{i}| \Theta_Z \rangle = C \, \sin \Theta_Z \,.
\ee
\ees
What is $C$ which appears in (\ref{thetaz})? The question concerning
the relationship between a heavy quark scalar condensate and the gluon condensate
has been studied long ago by \cite{svz}. There one has from the vanishing
of the VEV of the trace of the energy momentum tensor the following
relationship
\be
\label{scalar}
M_{Q}\, \langle \bar{Q}\,Q\,\rangle= -\langle \frac{\alpha_s}{12\,\pi}
\,G_{\mu \nu}^{a}\,G^{\mu \nu,a} \rangle \,.
\ee
In our case, the $\beta$ function is different. Instead of 
$-\frac{2}{3}\, n_{Q}$ in QCD, we have, because $\psi^{(Z)}_{i}$
belong to adjoint representations, the factor 
$-\frac{8}{3}\,n_{\psi^{(Z)}_{i}}$ in our case. As a result,
we have in our model
\be
\label{scalarz}
\sum_{i}\,m_{\psi^{(Z)}_{i}}\,\langle \bar{\psi}^{(Z)}_{i}\,
\psi^{(Z)}_{i} \rangle = -\frac{2}{3}\,
\langle\frac{\alpha_Z}{\pi}\,{\bf G}_{\mu \nu}^{(Z)}\,
{\bf G}^{\mu \nu,(Z)}\,\rangle \,.
\ee
Eq. (\ref{scalarz}) represents the case when $\Theta_Z = 0$. In consequence,
one can make the following identification
\be
\label{C}
C= -\frac{2}{3}\,
\langle\frac{\alpha_Z}{\pi}\,{\bf G}_{\mu \nu}^{(Z)}\,
{\bf G}^{\mu \nu,(Z)}\,\rangle \,.
\ee
(Note that one can also write 
$\langle \Theta_Z |\frac{2\,\alpha_Z}{\pi}\,{\bf G}_{\mu \nu}^{(Z)}\,
\tilde{{\bf G}}^{\mu \nu,(Z)}\, | \Theta_Z \rangle = 
-\frac{2}{3}\,
\langle\frac{\alpha_Z}{\pi}\,{\bf G}_{\mu \nu}^{(Z)}\,
{\bf G}^{\mu \nu,(Z)}\,\rangle \, \sin \Theta_Z$ which is
similar to the result obtained for QCD by \cite{halperin}.)
With (\ref{C}), we obtain
\be
\label{soft2}
V_B = \frac{4\,\pi}{3}\,\sin \Theta_Z \langle \frac{\alpha_Z}{\pi}\,
{\bf G}_{\mu \nu}^{(Z)}\,
{\bf G}^{\mu \nu,(Z)}\,\rangle \,(\frac{a_Z}{2\,\pi\,v_Z}) \,.
\ee
In a similar fashion to QCD sum rules \cite{svz}, we will assume that the
``$SU(2)_Z$ gluon'' condensate is given by the scale of strong interaction
as follows
\be
\label{gluecondense}
\langle \frac{\alpha_Z}{\pi}\,{\bf G}_{\mu \nu}^{(Z)}\,
{\bf G}^{\mu \nu,(Z)}\,\rangle \sim \Lambda_Z^4 \,. 
\ee
With this assumption, we finally obtain
\be
\label{soft3}
V_B = \frac{4\,\pi}{3}\,\sin \Theta_Z \,\Lambda_Z^4 \,
(\frac{a_Z}{2\,\pi\,v_Z}) \,.
\ee 

What are the implications of Eq. (\ref{soft3})? 
\bi

\item In the P and CP-conserving true vacuum with $\Theta_Z = 0$
($a_Z =0$), $V_B = 0$. The dark energy density
vanishes in this case.

\item The false vacuum will be one in which  $\Theta_Z \neq 0$ so
that $V_B \neq 0$ which implies a finite dark energy density.
The question is the following: What should $\Theta_Z$ be
so that the dark energy density is $\sim (2.4 \times 10^{-3}\,eV)^4$?
Notice that, in our model, the false vacuum is at $a_Z =
2\, \pi \, v_Z$.

\item We shall make the following proposal: The
false vacuum, where $a_Z = 2\, \pi \, v_Z$, is {\em maximally P and CP-violating} 
in the shadow
sector with
\be
\label{violation}
\Theta_Z = \frac{1}{4}(\frac{a_Z}{v_Z}) = \frac{\pi}{2} \,.
\ee
This implies that the dark energy density evaluated at $a_Z =
2\, \pi \, v_Z$ is
\be
\label{DE}
\rho_{vac} \sim ((\frac{4\,\pi}{3})^{1/4} \Lambda_Z )^4 \,.
\ee
This would imply that the scale where $SU(2)_Z$ grows strong is
approximately
\be
\label{lambdaz}
\Lambda_Z \approx (\frac{3}{4\,\pi})^{1/4} (2.4 \times 10^{-3}\,eV) 
\approx 1.7 \times 10^{-3}\,eV \,.
\ee
Such a scale was previously considered in \cite{hung1} and \cite{hung2}.

\ei

There is an important question to be addressed in our model or, for
that matter, in any model of dark energy: Are there {\em observable}
effects other than the acceleration of the present universe?
We have briefly touched upon that issue in \cite{hung2}. There
are however a few  additional points in this paper that we would
like to stress.
\bi

\item Our model contains in addition to particles
of the shadow factor, the messenger scalar field
$\tilde{\varphi}_{1}$ which carries quantum numbers of both 
the SM and the shadow sectors. The discovery of this particle
at colliders would provide an indirect pathway to the shadow
sector where most of the ``dark action'' takes place. 

\item The ``acceleron'' $a_Z$ is a bona-fide particle in our model
which could, in principle, communicate with the ``visible'' SM
sector through the messenger scalar field $\tilde{\varphi}_{1}$
which carries quantum numbers of both sectors. The following
issues is under investigation: 1) What are the possible particle effects
which can come from the the existence of $a_Z$?; 2) How feeble are they?

\item Finally, one might ask whether or not
there are other ``observable'' effects due to this
P and CP-violating vacuum. This issue is linked to the
above questions. Since $SU(2)_Z$ and its particles
represent the shadow sector (except for the messenger scalar field
$\tilde{\varphi}_{1}$), its effect
on the visible SM sector, in principle, would be tiny and hard to
detect. However, no matter how small such possible effects might be,
it would be interesting to find out what they are. This is
under investigation. (So far the only ``observable'' effect
of this vacuum in our model is the acceleration of the present universe.)

\ei

On another note, although the mechanism presented in this paper deals
only with the soft breaking term of the dark energy model of
\cite{hung2}, one might wonder if it is possible to apply the same idea
to a similar term discussed in \cite{sikivie} concerning the Peccei-Quinn
axion. Also the possibility of spontaneous P and CP violation in the
strong interactions has first been discussed by Dashen \cite{dashen}.
It was mentioned again in \cite{sikivie2} in the context of QCD. 
There was a renewed interest eight years ago in such a possibility, in particular
for hot QCD \cite{pisarski}, which could be looked for in
heavy ion collider experiments \cite{pisarski}. Earlier
works on ``obervable'' effects of QCD vacuum ``misalignment''
can be found in \cite{bj}. 
In this paper,
we propose the possibility that there exists such a false vacuum
but in the {\em shadow sector}.

In conclusion, we have presented a dynamical scenario in which
the size of the dark energy density $\rho_V \sim (10^{-3}\,eV)^4$
is related to a false vacuum which breaks P and CP due a 
non-vanishing condensate of shadow fermions and gauge bosons
of a new gauge group $SU(2)_Z$ which becomes strongly
interacting at a scale $\Lambda_Z \sim 10^{-3}\,eV$ \cite{hung2}.

\begin{acknowledgments}
This work is supported in parts by the US Department
of Energy under grant No. DE-A505-89ER40518. I would like
to thank the Aspen Center for Physics and the Spring Institute,
LNF, Frascati for hospitality while this work was completed. I also
would like to thank James D.(bj) Bjorken and Vernon Barger for useful
comments.
\end{acknowledgments}

\end{document}